\documentclass[aps,pra,twocolumn,twoside,superscriptaddress,notitlepage]{revtex4-2}
\usepackage{tikz}
\usetikzlibrary{calc}
\usetikzlibrary{shapes.geometric, arrows}
\usepackage{eufrak}
\usepackage{enumitem}
\usepackage{mathtools}
\usepackage{graphicx}
\usetikzlibrary{positioning}
\usepackage{bm}

\newcommand{\bematrix}{\left(\begin{matrix}}
\newcommand{\ematrix}{\end{matrix}\right)}
\usepackage{ulem} 
\usepackage[caption=false]{subfig}
\usepackage{dcolumn}
\usepackage{amsmath,amssymb}
\usepackage{bm}
\usepackage{bbm}
\usepackage{overpic}
\usepackage{latexsym}
\usepackage{color}
\usepackage[english]{babel}
\usepackage{latexsym}
\usepackage{psfrag,graphicx}
\usepackage{epsf}
\usepackage{amsmath}
\usepackage{amssymb}
\usepackage{amsfonts}
\usepackage{natbib}
\usepackage{multirow}
\usepackage{appendix}
\usepackage{verbatim}
\usepackage{enumitem}
\usepackage{dsfont}

\usepackage{float}
\usepackage{tikz}

\DeclareMathOperator{\Tr}{Tr}

\DeclareMathOperator{\QFT}{QFT}
\definecolor{mygrey}{gray}{0.35}
\definecolor{myblue}{rgb}{0.2,0.2,0.8}
\definecolor{myzard}{cmyk}{0,0,0.05,0}
\definecolor{mywhite}{rgb}{1,1,1}
\definecolor{myred}{rgb}{0.9,0.1,0.}
\usepackage[colorlinks=true,citecolor=myblue,linkcolor=myblue,urlcolor=myblue]{hyperref}

\usepackage[makeroom]{cancel}

\newenvironment{proof-of}[1]{\medskip\noindent\textbf{Proof of {#1}.}}{\hfill$\blacksquare$\medskip}
\newcommand{\ket}[1]{\left\vert#1\right\rangle}
\newcommand{\bra}[1]{\left\langle#1\right\vert}
\newcommand{\braket}[2]{\ensuremath{\langle #1 | #2 \rangle}}
\newcommand{\ketbra}[2]{\ensuremath{| #1 \rangle\!\langle #2 |}}

\usepackage{tcolorbox}

\begin{document}

\title{Nonlinear protocol for high-dimensional quantum teleportation}

\author{Luca Bianchi}
\affiliation{Department of Physics and Astronomy, University of Florence, 50019, Firenze, Italy}

\author{Carlo Marconi}
\affiliation{Istituto Nazionale di Ottica del Consiglio Nazionale delle Ricerche (CNR-INO), 50125 Firenze, Italy}

\author{Giulia Guarda}
\affiliation{European Laboratory for Non-Linear Spectroscopy (LENS), 50019 Sesto Fiorentino, Italy}
\affiliation{Istituto Nazionale di Ottica del Consiglio Nazionale delle Ricerche (CNR-INO), 50125 Firenze, Italy}

\author{Davide Bacco}
\affiliation{Department of Physics and Astronomy, University of Florence, 50019, Firenze, Italy}

\begin{abstract}
    \noindent Bell measurements, which allow entanglement between uncorrelated distant particles, play a central role in quantum communication. Indeed sharing, measuring and creating entanglement lie at the core of various protocols, such as entanglement swapping and quantum teleportation. While for optical qubit systems a Bell measurement can be implemented using only linear components, the same result is no longer true for high-dimensional states, where one has to consider either ancillary photons or nonlinear processes. Here, inspired by the latter approach, we propose a protocol for high-dimensional quantum teleportation based on nonlinear techniques. Moreover, we discuss the practical implementation of our proposed setup in the case of path-encoded qutrits, where nonlinear effects arise from sum-frequency generation. Finally, we compute the fidelity between quantum states to benchmark the validity of our model under the presence of crosstalk noise. Our approach is deterministic, scalable and does not rely on the use of auxiliary photons, thus paving the way towards the practical implementation of quantum networks based on nonlinear effects. \\ 
    
\end{abstract}

\date{September 29, 2024}

\maketitle


\section{Introduction}
The increasing advances in the field of quantum information are driving technological progress to an era where the realization of a quantum internet, i.e. a network of interconnected quantum devices, appears to be a feasible goal \cite{wehner2018quantum, ribezzo2023deploying, kimble2008quantum, pirandola2019end, pirandola2020advances}. In this context, it is of the utmost importance to devise a strategy to ensure effective connections between different devices within the same quantum network. 
A promising method to exchange data through distant nodes is to encode information in the quantum states of light. 
Indeed, photonic platforms are particularly convenient in this sense, since they allow to implement qubit systems using a variety of degrees of freedom (see, e.g. \cite{xu2020secure} and references therein).
Moreover, photons can be easily transmitted through optical fibers, a technology which is already used worldwide for classical communication and for the current internet network  \cite{bacco2021proposal}.
However, despite these advantages, the propagation of a quantum state is inevitably affected by detrimental phenomena, such as decoherence, dispersion, noise or attenuation, which hinder the possibility to realize a perfect transmission \cite{schlosshauer2004decoherence, pirandola2020advances, scarani2009security}. A common approach to mitigate some of these issues and to improve the capacity of the information carriers is to consider qudits, i.e. high-dimensional quantum states \cite{Dixon, Xiao, barreiro2008beating}. Information can be stored in photonic qudits in several ways 
\cite{cozzolino2019orbital,islam2017provably,kues2017chip,krenn2017entanglement,llewellyn2020chip, da2021path,zahidy2024practical}, but the transmission of quantum information between the nodes of a network still shows theoretical limitations \cite{pirandola2017fundamental}.
A common approach to circumvent this drawback is to use quantum repeaters \cite{azuma2023quantum, pirandola2015general}, i.e. devices that aim at transferring the entanglement between two distant non-interacting parties using an entanglement swapping protocol \cite{entanglementswapping}. In this context, a key challenge lies in the implementation of a Bell state measurement (BSM) \cite{braunstein1995measurement, pirandola2015advances}, i.e. a projection of a quantum state onto one of the Bell states, which represents a necessary step to entangle uncorrelated particles.

\noindent When restricting only to linear optical components, it has been theoretically proved that a BSM for qubit systems can be realized with an efficiency of $50\%$ \cite{lutkenhaus1999bell}, although such efficiency can be improved using auxiliary photons \cite{duvsek2001discrimination, olivo2018ancilla, ewert20143}. 
Another possibility to overcome this limitation is to release the assumption of linear optics and consider nonlinear processes \cite{kim2001quantum,zaidi2013beating}.
All of these rules are true for quantum systems limited to a two-dimensional Hilbert space, i.e. qubits. 
If we explore the case of qudit systems, a no-go theorem rules out the possibility to obtain a non-zero success probability in linear optics, without introducing ancillary photons \cite{carollo2001role, calsamiglia2002generalized}. Also in this case, nonlinear techniques have been recently considered to improve the efficiency of a teleportation protocol \cite{forbes}.

\noindent However, the solutions presented so far are not scalable and depend on the specific degrees of freedom chosen for the encoding, thus limiting the range of possible applications and demonstrations. Here, we extend the method originally presented in \cite{kim2001quantum} to the case of qudits, and propose a practical protocol to realize deterministic quantum teleportation for high-dimensional (HD) systems. Our approach does not require ancillary photons and is completely general, in the sense that it can be adapted to the case of a generic encoding and arbitrary dimension of the input state. 


\section{Qubit teleportation with nonlinear optics}
\label{sec2}
\begin{figure*}
    \centering
    \includegraphics[width=0.9\textwidth, height=11cm]{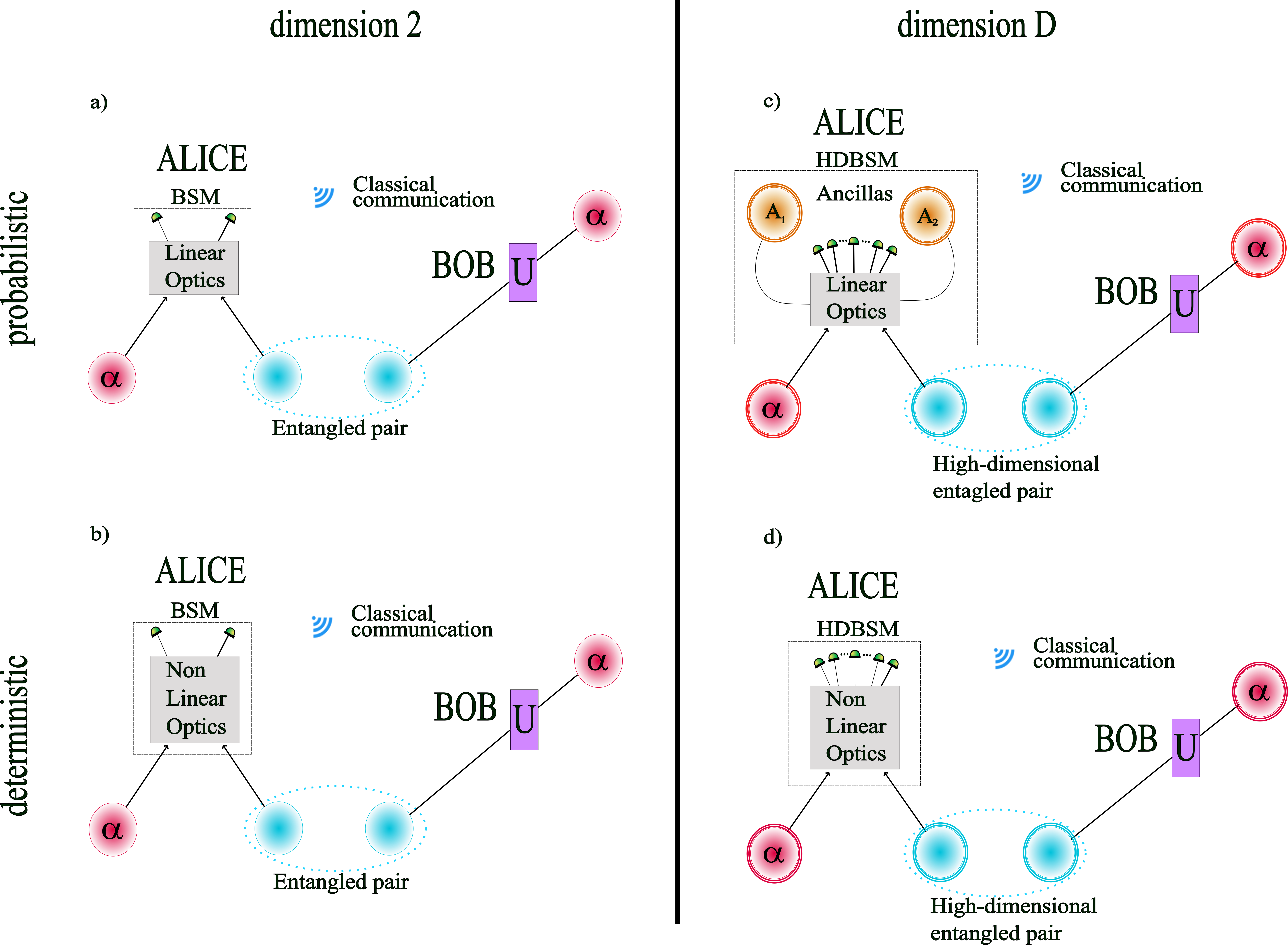}
    \caption{\textbf{Teleportation scheme: 2-dimensional vs. high-dimensional.} \textbf{a)} 2-dimensional teleportation scheme with a linear optical setup. This is the first teleportation protocol proposed and experimentally realized in quantum optics \cite{bouwmeester1997experimental}, whose main drawback is its intrinsically probabilistic nature. \textbf{b)} 2-dimensional teleportation scheme with nonlinear optics. Nonlinear optical implementation is advantageous as it provides a deterministic outcome \cite{kim2001quantum}. \textbf{c)} D-dimensional teleportation scheme with linear optics. Scaling up to higher dimensions requires the incorporation of ancillary states to perform a high-dimensional BSM (HDBSM) \cite{luo2019quantum}. \textbf{d)} D-dimensional teleportation scheme with nonlinear optics. This method allows for deterministic teleportation in higher dimensions, overcoming the probabilistic nature and the need for ancillary states of the measurement.
    }
    \label{teleportation_scheme}
\end{figure*}

In the standard teleportation protocol \cite{{bennett1993teleporting}}, Alice, the sender, wants to transfer an unknown quantum state, $\ket{\phi}_{A_1}$, to Bob, the receiver. In order to successfully perform this operation, it is necessary that Alice shares with Bob a maximally entangled state $\ket{\Psi}_{A_2 B}$. Then, Alice performs a BSM on her particles in order to project her quantum state onto one of the Bell states \cite{einstein1935can}. After the exchange of classical communication, Bob applies a unitary transformation, corresponding to one of the matrices of the $SU(2)$ group, i.e. $\{\mathds{1},\sigma_x, i\sigma_y , \sigma_z\}$ \cite{sakurai2020modern}, to rotate the quantum state and recover the state $\ket{\phi}_{B}$ (see Fig.\ref{teleportation_scheme}a). When representing this protocol as a quantum circuit, the BSM performed by Alice becomes a CNOT gate, followed by a Hadamard gate on her particles \cite{nielsen2001quantum}. 
The experimental implementation of a CNOT represents a major challenge in quantum optics due to the fact that the energies of the photons are not sufficient to trigger photon-photon interactions \cite{baur1988electromagnetic}. As a consequence, when restricting to linear optical devices, a CNOT gate cannot be deterministically realized, although there exist probabilistic implementations, based on auxiliary photons and conditional measurements, such as the KLM method \cite{knill2001scheme}. To circumvent this issue, a common approach is to exploit the indistinguishability of photons impinging on a beam splitter to leverage the Hong-Ou-Mandel effect \cite{hong1987measurement}. The correlated output pair is then measured by photodetectors whose clicks are associated with Bell states \cite{bouwmeester1997experimental}. However, protocols based on this technique, are still probabilistic \cite{lutkenhaus1999bell}, and not scalable to qudits unless auxiliary photons are considered \cite{calsamiglia2002generalized} (see Fig.\ref{teleportation_scheme}c). 
Moving to nonlinear optical techniques, a variety of phenomena involving light-matter coupling can be considered \cite{boyd2008nonlinear}. Among these, SFG describes how two input photons, entering a non-linear crystal characterized by a $\chi^{(2)} $ dielectric susceptibility, can be annihilated and converted into a new highly-energetic photon.
The interacting Hamiltonian that describes this process is given by
\begin{equation*}
 H_{int} \propto \hbar \chi^{(2)} (\hat{c}\hat{a}^{\dagger} \hat{b}^{\dagger} - \hat{c}^{\dagger}\hat{a}\hat{b} )~,
\end{equation*}
where $\hat{a}$ and $\hat{b}$ are the annihilation operators of the input modes and $\hat{c}$ is the one of the output mode. Here, $\chi^{(2)}$ is a susceptibility constant which depends on the timescale of the interaction, the length of the crystal and the effective refractive index of the medium. In order for the process to be efficient, the frequencies of the optical modes should satisfy the following relation, i.e. $\omega_c = \omega_a + \omega_b$ \cite{mandel1995optical}, which reflects the so called phase matching condition $\vec{k}_c \simeq \vec{k}_a + \vec{k}_b$. To increase the efficiently of the SFG process, one of the two input modes can be designed to be a strong pump pulse.

\noindent In the NLO scenario (see Fig.\ref{teleportation_scheme}b) presented in \cite{kim2001quantum}, the linearly polarized input qubits, $A_1$ and $A_2$, impinge on two groups of nonlinear $\chi^{(2)}$ crystals, which replace the action of a CNOT gate. In this scheme, the state $A_1$ to be teleported is used as a pump to trigger the SFG process. If the input polarizations match those required by either one of the crystals, the corresponding photons undergo SFG, resulting in an upconverted photon $C$. Then, using a dichroic mirror, the upconverted photon is sent to a polarizing beam splitter. Finally, a set of wave-plates rotates the polarization into the left-right basis $(\ket{L}, \ket{R})$ and the output is measured. Depending on the measurement outcome, suitable rotation matrices must be applied in order to retrieve the original state to be teleported. Notice that, if the two incoming photons are not converted by the first couple of type I crystals, they will be inevitably converted by the second couple of type II crystals. Hence, the four possible combinations of polarization outcomes for photon $C$ ($\ket{L}$ or $\ket{R}$) and crystal type triggered by photons $A_1$ and $A_2$ (type I or type II) are in one-to-one correspondence with the four Bell states of systems $A_1 A_2$ (see Table \ref{tab:multirow-cline}). In fact, for each Bell state, there exists a corresponding pair (type of crystal, polarization outcome) that can be converted into the original state to be teleported, using the same matrix rotation.

\noindent For the sake of clarity, we present a concrete example of the NLO teleportation. The logical states of the qubit are encoded in polarization, i.e. $\ket{0} \coloneq \ket{H}, \ket{1} \coloneq \ket{V}$. The input state A to be teleported is then found in the state
\begin{equation*}
\ket{\phi}_{A_1} = \alpha \ket{0}_{A_1} + \beta \ket{1}_{A_1}~.
\end{equation*}
The entangled pair shared between Alice and Bob is generated, e.g. by an spontaneous-parametric-down-conversion (SPDC) source, and is given by
\begin{equation*}
\ket{\Psi}_{A_2 B} = \frac{1}{\sqrt{2}}(\ket{00}_{A_2 B}+\ket{11}_{A_2 B})~.
\end{equation*}
Therefore, the total state of the system can be rewritten as
\begin{align*}
    \ket{\Psi}_{A_1 A_2 B} = &\frac{1}{2}[(\alpha \ket{00}_{A_1 A_2 }+ \beta\ket{10}_{A_1 A_2 })\ket{0}_B \\
    & +(\alpha \ket{01}_{A_1 A_2}+ \beta\ket{11}_{A_1 A_2})\ket{1}_B ]~.
\end{align*}
Type I crystals convert photons with the following rules:
\begin{equation}\begin{split}
        &\ket{HH} \rightarrow \ket{V}~,\\
        &\ket{VV} \rightarrow \ket{H}~,
\end{split}\end{equation}
and analogous relations hold for type II crystals, i.e.
\begin{equation}\begin{split}
        &\ket{HV} \rightarrow \ket{V}~, \\
        &\ket{VH} \rightarrow \ket{H}~.
\end{split}\end{equation}
These rules can also be expressed through non-squared matrices  $M^{(2)}_{i} : \mathcal{H}_{A_1} \otimes \mathcal{H}_{A_2} \rightarrow \mathcal{H}_{B}$ with $i=0,1$, where the superscript reminds that $\dim(\mathcal{H}_{X})=2,~\forall X \in \{A_1,A_2,B\}$. Such operators can be explicitly represented as follows, i.e.
\begin{equation}
\label{povm2}
    M^{(2)}_0 = \begin{pmatrix}
        0 & 0 & 0 & 1 \\
        1 & 0 & 0 & 0 
    \end{pmatrix}~, \qquad
    M^{(2)}_1 = \begin{pmatrix}
        0 & 1 & 0 & 0 \\
        0 & 0 & 1 & 0
    \end{pmatrix}~.
\end{equation}
For instance, in the case of two input photons triggering a type I crystal, the joint state is given by
\begin{equation*}
    \ket{\Psi}_{CB} = \frac{1}{\sqrt{2}}(\alpha \ket{10}_{CB} + \beta \ket{01}_{CB})~,
\end{equation*}
where $C$ is the state of the upconverted photon.
Due to the phase matching conditions, photon $C$ is created with an energy $\omega_C \simeq \omega_{A_{1}} +\omega_{A_{2}}$ and it is therefore reflected by the dichroic mirror and measured in the diagonal basis. Measuring the photons in the basis $\ket{L}, \ket{R}$ yields
\begin{equation}
    \ket{\Psi}_{CB} = \frac{1}{2}[\ket{L}_C (\alpha \ket{0}_B + \beta \ket{1}_B) + \ket{R}_C (\beta \ket{0}_B - \alpha \ket{1}_B)]~.
\end{equation}
If photon $C$ is found to be in the state $\ket{L}$, Bob's qubit state becomes
\begin{equation*}
\ket{\psi}_B = \alpha \ket{0}_B + \beta \ket{1}_B~,
\end{equation*}
and no rotation should be applied. On the other hand, if the upconverted photon comes from another set of crystals or is found in the state $\ket{R}$, a unitary transformation is required to convert $\ket{\phi}_B$ into the desired initial state. The rules for such rotations can be summarized as in Table \ref{tab:multirow-cline}.
\begin{table}[h!]
    \centering
    \begin{tabular}{|c|c|c|}
        \hline
        Crystal type & Polarization & C. C. matrix \\
        \hline
        \multirow{2}{*}{Type I} 
        & $\ket{L}$ & $\mathds{1}$ \\
        & $\ket{R}$ & $\sigma_z$ \\ 
         \hline
       \multirow{2}{*}{Type II} 
        & $\ket{L}$ & $\sigma_x$  \\
        & $\ket{R}$ & $i\sigma_y$  \\ 
        \hline
    \end{tabular}
    \caption{Classical communication matrices associated to crystal groups and output polarization measurements of the upconverted photon.}
    \label{tab:multirow-cline}
\end{table}

\noindent To summarize, the uncorrelated pair $A_1A_2$ enters the type I and type II sets of crystals. The upconverted photon $C$ is measured in the $(\ket{L},\ket{R})$-polarization basis. This procedure is equivalent to project the uncorrelated photons onto one of the Bell pairs. The measurement outcome is then communicated to the receiver through a classical channel, and conditionally to this result, a unitary operation is applied to qubit $B$. Before introducing our results, it is important to stress that the protocol presented here allow to implement deterministic quantum teleportation. Indeed, once the SFG process is triggered, Bell state discrimination can be realized with certainty, as opposed to protocols relying on linear optical devices.

\section{Results}
\subsection{Generalization to qudit teleportation}
\label{sec3a}
We now extend the teleportation protocol introduced in the previous section to the case of arbitrary high-dimensional systems (see Fig.\ref{teleportation_scheme}d). To this end, we first recall how the quantum teleportation protocol can be recast using the formalism of positive operator-valued measurements (POVMs) . 
This approach is particularly convenient to our analysis since it allows to extend the action of the nonlinear crystals, described by Eq.(\ref{povm2}), to the case of qudit systems. Second, we present a generalization of the previous NLO protocol, together with a proposal for its experimental implementation. The formalism we develop is entirely general, as it can be adapted, in principle, to the case of alternative encodings as well as other nonlinear techniques.

\noindent In the case of bipartite $d$-dimensional systems, Bell states are given by \cite{sych2009complete}
\begin{equation}
\label{HD_Bell}
    \ket{\Psi_{lm}} = \sum_{k=0}^{d-1}\omega_{d}^{lk}\ket{k}\ket{k \oplus m}~,
\end{equation}
where $l,m ={0,\dots,d-1}$ denote the logical levels of the qudits, $\omega_d = e^{2\pi i / d}$ is the $d$-th root of the unity, and $\oplus$ denotes the sum modulo $d$. In the most general scenario, the teleportation protocol can be equivalently recast resorting to the formalism of density operators. 
In this case, the initial state of the overall system can be written as
\begin{equation}
    \rho_{A_1 A_2 B } = \rho_{A_1} \otimes \rho_{A_2 B}~,
\end{equation}
where $\rho_{A_1} = \ketbra{\phi}{\phi}$ is the state to be teleported and $\rho_{A_2 B} = \ketbra{\Psi_{00}}{\Psi_{00}}$ is the Bell state of Eq.(\ref{HD_Bell}) with $l=m=0$. Here we assume $\ket{\phi} = \sum_{i=0}^{d-1} c_{i} \ket{i}$, where the complex coefficients $c_i$ satisfy the normalization constraint $\sum_{i=0}^{d-1} |c_{i}|^{2} = 1$. Hence, Alice performs a measurement on her qudits $A_1, A_2$ described by a POVM, i.e. a collection of operators $\{\Pi^{(d))}_{m}\}$ such that
\begin{equation}
    \Pi^{(d)}_{m} =(M^{(d)}_{m})^{\dagger} M^{(d)}_{m} \succeq 0~, \quad \sum_{m} \Pi^{(d)}_{m}= \mathds{1}_{A_1 A_2},
    \label{POVM}
\end{equation}
where $M^{(d)}_{m}$ are measurement operators and $m$ labels the measurement outcomes. Conditional to each outcome, $\rho_{A_1 A_2 B}$ collapses to the state 
\begin{equation}
\rho^{(m)}_{A_1 A_2 B} = \left(M^{(d)}_m \otimes \mathds{1}_{B}\right) \rho_{A_1 A_2 B}  \left( (M^{(d)}_{m})^{\dagger} \otimes \mathds{1}_{B}\right)~,
\end{equation}
with probability $p_{m} = \Tr_{A_1 A_2 B} \{(M^{(d)}_{m})^{\dagger} M^{(d)}_{m} \rho_{A_1 A_2 B} \}$.
Depending on Alice's measurement outcome, Bob applies a suitable rotation to its share of the state $\rho^{(m)}_{A_1 A_2 B}$, thus concluding the protocol.

\noindent When dealing with the NLO scenario presented in the previous section, the action of the nonlinear crystals of Eqs.(\ref{povm2}) can be generalized to the qudit case introducing the following matrices, i.e.
\begin{equation}
    \begin{split}
    &M^{(d)}_{0} = \sum_{k=0}^{d-1}\ketbra{k\oplus 1}{k}\bra{d \ominus k}~,\\
    &M^{(d)}_{1} = \sum_{k=0}^{d-1}\ketbra{k}{k}\bra{d \ominus (k+1)}~,\\
    &M^{(d)}_{j} = \sum_{k=0}^{d-1}\ketbra{k\oplus j}{k}\bra{d \ominus (k+j)}~, \quad j=2,\dots,d-1~,
    \end{split}
    \label{misure}
\end{equation}
where $\oplus$ and $\ominus$ denote the addition and subtraction modulo $d$, respectively. It is straightforward to check that the operators $\frac{1}{\sqrt{d}}M^{(d)}_{m}$ satisfy the relation
\begin{equation}
    \frac{1}{d}\sum_{m=0}^{d-1} M^{(d) \dagger}_m M^{(d)}_{m} = \mathds{1}~,
    \label{compl}
\end{equation}
where the normalization factor accounts for the probability of getting the outcome $m$. 

\noindent In the HD case, a unitary transformation between two orthonormal basis can be realized by the quantum Fourier transform (QFT) \cite{hoyer1997efficient}, formally defined for $d$-dimensional systems as 
\begin{equation}
\label{qft}
    \QFT_d = \frac{1}{\sqrt{d}}\begin{pmatrix}
        1 & 1 & 1 & \dots & 1 \\
        1 & \omega_d & \omega_d^2 & \dots & \omega_d^{d-1} \\
        \vdots & \vdots & \vdots & \dots &  \vdots\\
        1 & \omega_d^{d-1} & \omega_d^{2(d-1)} & \dots & \omega_d^{(d-1)^2}~.
    \end{pmatrix}~,
\end{equation}
\noindent whose action on a generic vector $\ket{x} \in \mathbb{C}^{d}$ is given by
by \cite{nielsen2001quantum}
\begin{equation}
 \ket{x} \rightarrow \frac{1}{\sqrt{d}}\sum_{k=0}^{d-1} \omega^{xy}_{d}\ket{y}~.
\end{equation}
The last step of the protocol is given by photodetection measurement, that projects a photon into one of the states $\{\ket{i}\}_{i=0}^{d-1}$ of the computational basis. Thus, Alice can construct measurement operators of the form:
\begin{equation}
    M_{(i, m)} = \ketbra{i}{i}\QFT_d M^{(d)}_m ~,
\end{equation}
corresponding to the following set of POVMS, $\{\Pi_{(i,m)}\}$, with
\begin{equation}
    \Pi_{(i,m)} = M_{(i, m)}^{\dagger}M_{(i, m)}~, \quad i, m = 0,\dots,d-1.
\end{equation}
Notice that, due to their explicit expression, matrices $ \Pi_{(i, m)}$ are positive by construction. Moreover, using the unitarity of the operator $\QFT_d$, it is straightforward to check that $ \Pi_{(i, m)}$ verify the completeness relation, certifying them as a POVM set. 
As in the two-dimensional case, Alice broadcasts her results $(i,m)$ to Bob, who conditionally applies an HD version of classical communication rotations. Such transformations can be conveniently expressed by means of Weyl matrices as \cite{dutta2023qudit}:
\begin{equation}
\label{CC}
    U_{im} = \sum_{k=0}^{d-1}\omega_{ki}\ketbra{k}{k \oplus m }~.
\end{equation}
We emphasize that $i$ and $m$ denote the outcome of the photodetection and the triggered nonlinear crystal, respectively.

\subsection{Experimental proposal}
\label{sec3b}

\noindent In this section, we propose an experimental implementation of our HD protocol using path-encoded qudits. In this scheme, 
we exploit $d^2$ identical type-0 $\chi^{(2)}$ crystals to generate an upconverted photon for each possible combination of input paths. Stated differently, for each pair of paths associated to the two input photons, there exists only one triggered crystal. As a consequence, by performing a measurement of the output photons paths, it is possible to implement the analogous of a HDBSM.
\begin{figure*}
    \centering
    \includegraphics[width=0.9\textwidth, height=9.8cm]{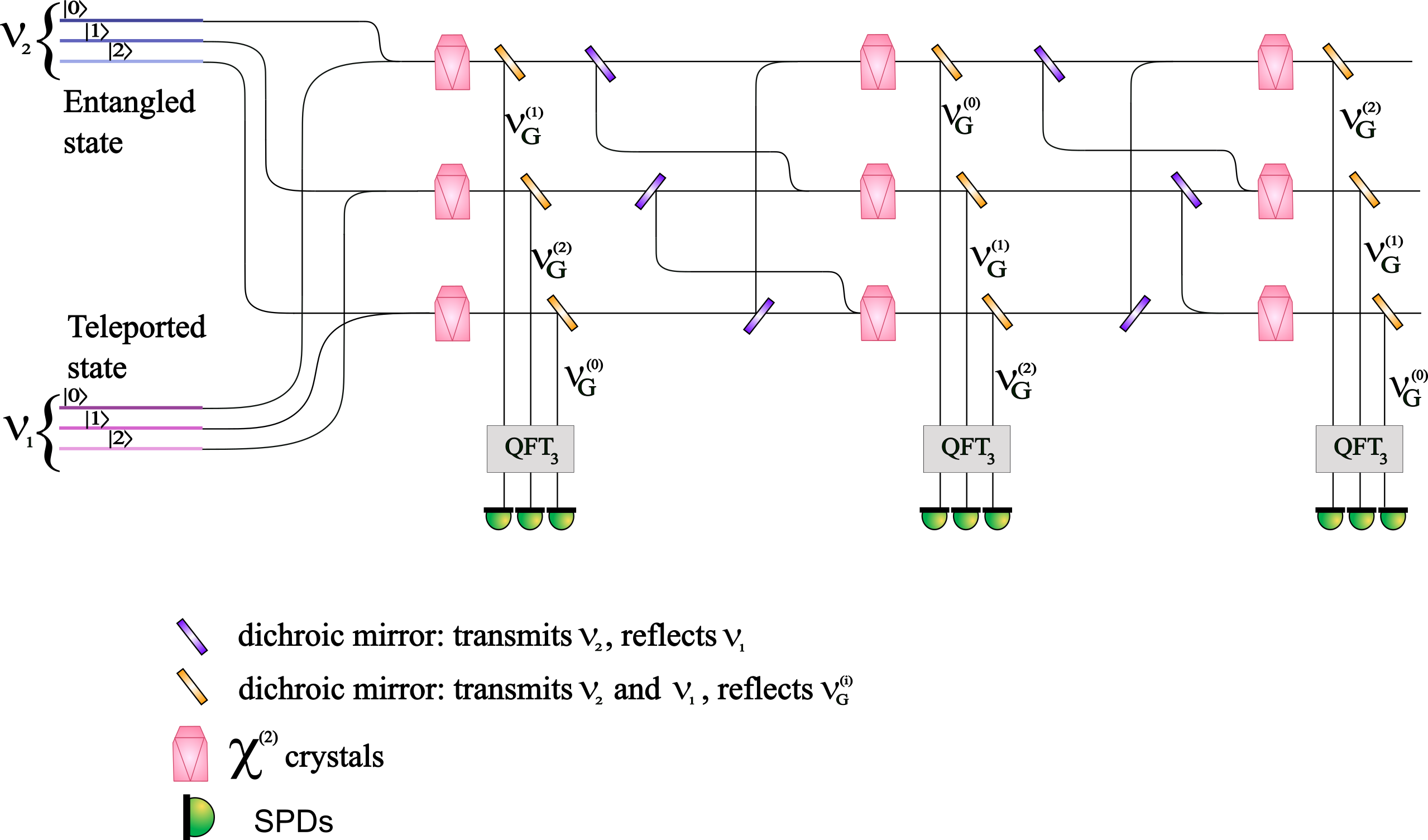}
    \caption{\textbf{Nonlinear optical measurement for quantum teleportation of a qutrit:} the two qutrits $A_1, A_2$ owned by Alice are path-encoded photons. $A_1$ represents the qutrit to be teleported, while $A_2$ comes from an entangled pair shared by Alice and Bob. Each pair of paths enters a set of type-0 $\chi^{(2)}$ crystals according to Eq.(\ref{misure}). If the upconversion takes place, the resulting photon will be reflected by the first set of dichroic mirrors (yellow). On the other hand, if the conversion does not happen, a second set of dichroic mirrors (purple) redistributes the photons into the other groups of crystals. Once the SFG process is triggered, the new photon $C$ is in one of the three paths and undergoes a $\QFT_3$ transformation that mixes its modes. Finally, a set of three SPDs detects the photon into one of the possible paths. Alice then broadcasts the information about the measured path and the set of triggered crystals to Bob, who conditionally applies a unitary 3D-Weyl transformation to recover the initial superposition into its qutrit $B$ (see Eq.(\ref{CC})).}
    \label{experimental_setup}
\end{figure*}
For the sake of simplicity, let us now stick to the qutrit case ($d=3$). The details of the setup are shown in Fig.\ref{experimental_setup}. Our goal is to teleport an unknown quantum state $\rho_{A_1}$. Such state is given by a photon at a frequency $\nu_1$ that stores information on three different paths. An entangled pair of photons, $\rho_{A_2 B}$, is  generated through SPDC process \cite{guerreiro2014nonlinear}. Here a 532 $nm$ laser pumps a periodically poled lithium niobate (PPLN) nonlinear crystal, quasi-phasematched to generate an entangled photon pair at $\nu_2 = 1560\ nm$ ($\ket{\phi_{A_2}}$) and $\nu_3 = 807\ nm$ ($\ket{\phi_{B}}$) \cite{guerreiro2014nonlinear}. The frequency of the unknown quantum state $\ket{\phi_{A_1}}$, namely $\nu_1$, must be different from the one of the signal $A_2$. This is a crucial assumption to distinguish the photons at the exit of a given set of crystals, whenever the SFG process is not triggered. Following the experimental implementation of a SFG with two single photons \cite{guerreiro2014nonlinear, guerreiro2013interaction}, we set $\nu_1 = 1551\ nm$. To trigger the type-0 crystal, the two photons $A_1$ and $A_2$ must also have the same polarization. The paths of the input photons of our measurement scheme enter the crystals according to the rules given in Eq.(\ref{misure}). For qutrits, the set of crystal transformations read
\begin{equation}
    \begin{split}
        M_0^{(3)} = \ketbra{0}{21}+\ketbra{1}{00}+\ketbra{2}{12}~, \\
        M_1^{(3)} = \ketbra{0}{01}+\ketbra{1}{10}+\ketbra{2}{22}~,\\
        M_2^{(3)} = \ketbra{0}{20}+\ketbra{1}{11}+\ketbra{2}{02}~.
    \end{split}
\end{equation}
All of the nine crystals are identical type-0 PPLN \cite{parameswaran2002highly} and the quasi-phase matching condition allows for the conversion of the two single photons $A_1$ and $A_2$ into a photon $C$ at frequency $\nu_G = 810\ nm$. SFG with single photons as input states has already been investigated \cite{guerreiro2014nonlinear, guerreiro2013interaction} and an experimental efficiency $\eta = (1.5 \pm 0.3) \times 10^{-8}$ was measured \cite{guerreiro2014nonlinear}. After each crystal, a dichroic mirror reflects the upconverted photon and transmits the input ones. A second mirror is then placed to reflect photons with frequency $\nu_1$ and transmit the ones with frequency $\nu_2$. The separation of the frequencies allows to recombine the paths and include of all the possibilities described by the measurement operators. If one of the set of crystals is triggered, the upconverted photon takes one of the three possible paths at the exit of the nonlinear media. It is worth noting that the new photon has the same frequency $\nu_G$ independently on which crystal was triggered. The paths are then recombined and mixed according to a $\QFT_3$ transformation (see Eq.(\ref{qft}) with $d = 3$), that can be realized with a combination of beam splitters and phase shifters. Finally, three single photon detectors (SDPs \cite{lunghi2012advantages}) measure the presence of the upconverted photon in one of the three paths. Then Alice communicates through a classical channel her results (measured path in the computational basis and group of triggered crystals) to Bob, that in accordance to this information rotates his own qutrit. A practical implementation of such rotations for path encoded qutrits can be found in \cite{luo2019quantum}.

\subsection{Fidelity evaluation}
\label{sec3c}
\begin{figure}[H]
    \centering
    \includegraphics[width=0.5\textwidth]{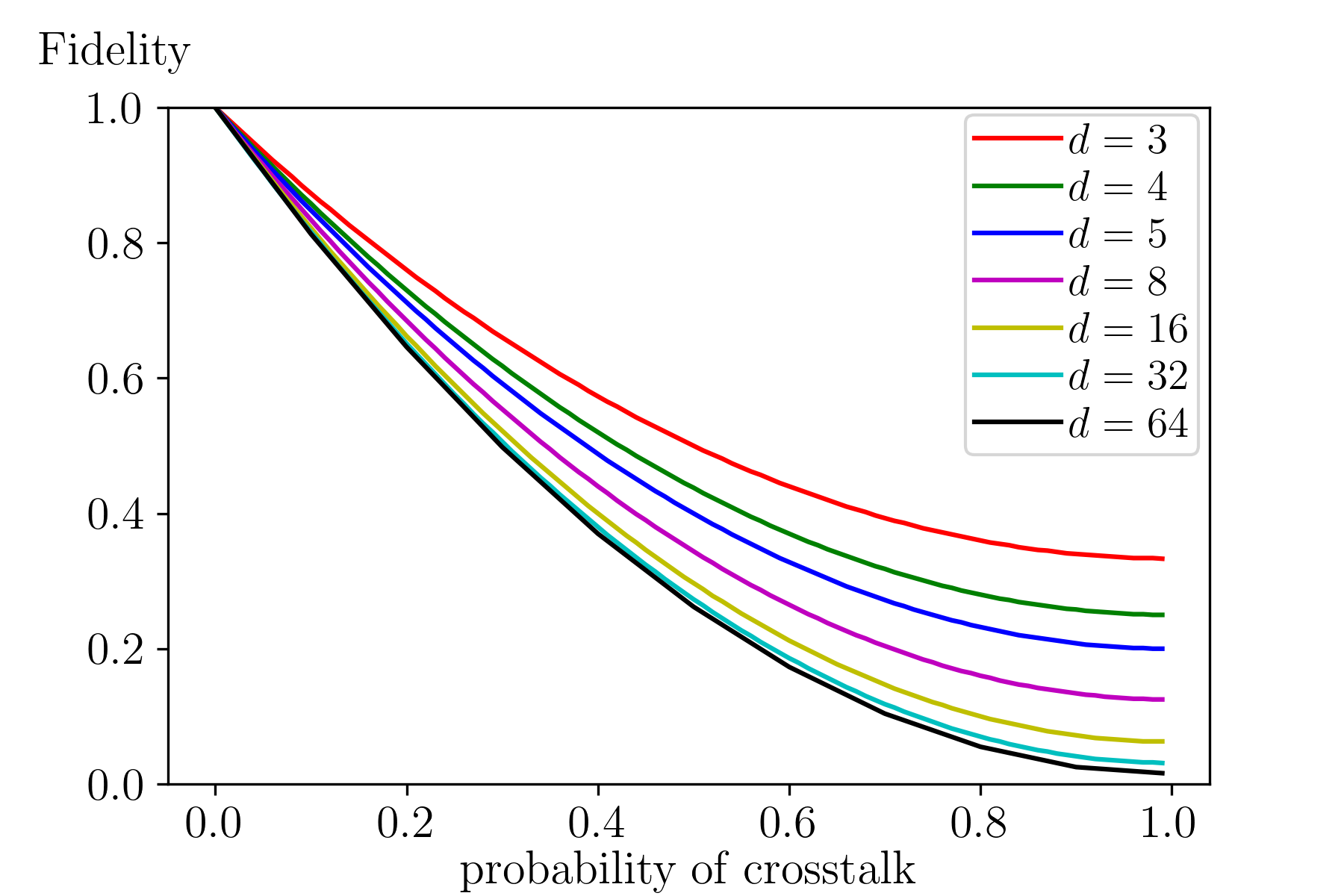}
    \caption{Comparison of the fidelity between an input state $\ket{\phi}_{A_1} = \frac{1}{\sqrt{d}}\sum_{k=0}^{d-1}\ket{k}$ and the output state of our protocol, for different values of the dimension $d$. Here, $p$ quantifies the effect of the crosstalk noise (see main text for further details). The efficiency of the photon conversion is supposed to be $1$.}
    \label{graficone}
\end{figure}

To certify the HD NLO teleportation scheme, we simulate the protocol proposed in the previous section for different dimensions. In particular, we consider a noisy scenario to account for realistic deviations from the ideal behaviour and compute the fidelity between the output state and the initial state to be teleported. 
To this end, we model the effect of noise as a quantum operation $\mathcal{E}$ whose action on a generic quantum state $\rho$ is expressed by \cite{wolf2012quantum} 
\begin{equation}
\label{channel}
    \mathcal{E}(\rho) = \sum_{i=1}^{r}C_i \rho C_i^{\dagger}~, \quad r\leq d^{2}~,
\end{equation}
where $C_{i}$ are Kraus operators, satisfying $\sum_{i}C_i^{\dagger}C_i = \mathds{1}$.
Moreover, we assume the noise to act on each qudit independently, so that the noise acting on Alice's qudits can be assumed to be the tensor product of two independent sets of Kraus operators \cite{nielsen2001quantum}. On the other hand, we suppose Bob's qudit is not affected by any type of noise, since his share of the entangled pair is distributed before performing the upconversion and the subsequent measurement processes. Under this assumption, each Kraus operator in Eq.(\ref{channel}) can be factorized as
\begin{equation*}
    C_{i} =C_{i, A_1} \otimes C_{i, A_2} \otimes \mathds{1}_{B}~.
\end{equation*}
To assess the performances of our model, we focus on the so-called crosstalk noise, which represents one of the main sources of loss in path-encoded quantum states \cite{dualrail}. The effect of such crosstalk can be formally modeled by a set of Kraus operators acting as a $d$-flip channel, i.e. \cite{dutta2023qudit}
\begin{equation}
    \begin{cases}
        C_0 = \sqrt{1-\frac{(d-1)p}{d}} \mathds{1}~, \\
        C_i = \sqrt{\frac{p}{d}}U_{0i}, \quad i = 1 \dots d-1~,
    \end{cases}
\end{equation}
where $U_{0i}$ are the operators of Eq.(\ref{CC}) and $p$ is the probability of flipping the input state into one of the other $d-1$ levels. Since the protocol is independent on the encoding as well as the NLO process employed, we stick to the case where the efficiency of the NLO conversion is ideally $1$. We also assume as a figure of merit of our protocol the fidelity between two quantum states, formally defined as \cite{jozsa1994fidelity}
\begin{equation}
    F(\rho,\sigma) = \Tr\{\sqrt{\sqrt{\rho} \sigma\sqrt{\rho}}\}~,
\end{equation}
where $\rho$ and $\sigma$ are two generic quantum states.
In the case of an initial pure state $\rho = \ketbra{\phi}{\phi}$, it can be shown that the fidelity can be reduced to the much simpler expression $ F(\rho,\sigma)= \sqrt{\braket{\phi}{\sigma|\phi}}$ \cite{nielsen2001quantum}.
Our proposed setup has been simulated for dimensions $d = \{2,3,4,5,8,16,32,64\}$, assuming that qudit $A_1$ is prepared in the state $\ket{\phi}_{A_1} = \frac{1}{\sqrt{d}}\sum_{k=0}^{d-1}\ket{k}$, while the Bell pair is given by the state $\ket{\Psi_{00}}_{A_2 B}$. As it can be observed in Fig.\ref{graficone}, the fidelity of the teleported superposition decreases as the crosstalk probability grows. Furthermore, as the dimension of the system increases, the fidelity decays more rapidly, a behavior which is in good agreement with the results reported in \cite{forbes, fonseca2019high, jankovic2024noisy}, where the effect of several classes of noise on the performances of HD quantum teleportation was investigated. 

\section{Conclusions}
\noindent In this work, we have designed a scheme for deterministic high-dimensional quantum teleportation based on nonlinear optics. In order to illustrate our method, we consider path-encoded qudits and assume sum-frequency generation as the interacting process between photons. In addition, our approach is general and can be applied also to different encodings and nonlinear techniques. Moreover, our proposed protocol is deterministic and does not require the use of ancillary photons, being also scalable up to arbitrarily high-dimensions. To demonstrate the feasibility of our approach, we propose an experimental setup for path-encoded qutrits. Finally, we validate the effectiveness of our setup by computing the fidelity of the teleported qudit in the presence of external noise for different dimensions. Our results show a good agreement with those previously reported in the literature, thus offering a robust framework for implementing high-dimensional quantum teleportation with nonlinear optics. We believe our findings will open new avenues of research towards the practical implementation of a quantum network.

\section{Acknowledgements}
This research has been cofunded by the European Union ERC StG, QOMUNE, 101077917, and by the NextGeneration EU, "Integrated infrastructure initiative in Photonic and Quantum Sciences" - I-PHOQS [IR0000016, ID D2B8D520, CUP B53C22001750006].

\bibliography{biblio}

\end{document}